# On the orbital variability of Ganymede's atmosphere

F. Leblanc[1], A.V. Oza[1], L. Leclercq [2,3], C. Schmidt[1], T. Cassidy [4], R. Modolo.[2], J.Y. Chaufray,[2] and R.E. Johnson[3]

[1] LATMOS/IPSL, UPMC Univ. Paris 06 Sorbonne Universités, UVSQ, CNRS, 4 place Jussieu 75005 Paris, France

[2] LATMOS/IPSL, UVSQ Université Paris-Saclay, UPMC Univ. Paris 06, CNRS, Guyancourt, France

[3] University of Virginia, Charlottesville, Virginia, USA

[4] Laboratory for Atmospheric and Space Physics, University of Colorado, Boulder, CO 80303, USA

**Abstract :** Ganymede's atmosphere is produced by radiative interactions with its surface, sourced by the Sun and the Jovian plasma. The sputtered and thermally desorbed molecules are tracked in our Exospheric Global Model (EGM), a 3-D parallelized collisional model. This program was developed to reconstruct the formation of the upper atmosphere/exosphere of planetary bodies interacting with solar photon flux and magnetospheric and/or the solar wind plasmas. Here, we describe the spatial distribution of the $H_2O$ and $O_2$ components of Ganymede's atmosphere, and their temporal variability along Ganymede's rotation around Jupiter. In particular, we show that Ganymede's $O_2$ atmosphere is characterized by time scales of the order of Ganymede's rotational period with Jupiter's gravity being a significant driver of the spatial distribution of the heaviest exospheric components. Both the sourcing and the Jovian gravity are needed to explain some of the characteristics of the observed aurora emissions. As an example, the $O_2$ exosphere should peak at the equator with systematic maximum at the dusk equator terminator. The sputtering rate of the $H_2O$ exosphere should be maximum on the leading hemisphere because of the shape of the open/close field lines boundary and displays some significant variability with longitude.

**I Introduction**

Ganymede is the only known satellite with an intrinsic magnetosphere (Kivelson et al. 1996). Its icy surface is eroded by the solar radiation flux and by the Jovian energetic plasma leading to a locally-collisional atmosphere and a locally surface-bounded exosphere (Stern 1999; Marconi 2007; Turc et al. 2014; Plainaki et al. 2015; Shematovich 2016). Unfortunately, Ganymede's atmosphere remains essentially unknown despite the flybys during the mission Galileo in the 90s. This mission detected only one atmospheric emission, Lyman α, interpreted as being produced by resonant scattering by exospheric atomic hydrogen (Barth et al. 1997) with a column density around $9.21 \times 10^{12}$ H/cm$^2$. Such a column density and its variation with distance to Ganymede suggested a surface density around $1.5 \times 10^4$ H/cm$^3$ and an exospheric temperature around 450 K. Hubble Space Telescope was also able to identify another component of Ganymede atmosphere, the $O_2$ molecule through the observations of two UV atomic oxygen emissions with intensities corresponding to a column density around 1 to $10 \times 10^{14}$ $O_2$/cm$^2$ (Hall et al. 1998). Since this discovery, several other HST observations of these emission have revealed their auroral origin (Feldman et al. 2000; McGrath et al. 2013; Saur et al. 2015) highlighting the relations between the spatial distribution of these emissions, Ganymede's intrinsic magnetospheric structure (Kivelson et al. 1996) and Ganymede's surface's albedo (Khurana et al. 2007).

Gurnett et al. (1996) published the first observations of Ganymede's ionosphere by the plasma wave spectrometer (PWS) on board Galileo suggesting ion densities on the order of 100 cm$^{-3}$ and a scale height around 1000 km. This first set of measurements was later complemented by a second flyby (Eviatar et al. 2001b) confirming the density range but suggesting that a number density on the order of 1000 cm$^{-3}$ at the surface may be more realistic and in better agreement with the upper limit set by radio occultation (Kliore 1998). Measurements by the Plasma Science (PLS) instruments were also performed during Galileo flybys in the polar regions of Ganymede (Frank et al. 1997). These authors reported the observations of hydrogen ions outflowing from Ganymede's poles with temperatures around $4 \times 10^4$ K and surface densities around 100 cm$^{-3}$. These observations were later reanalyzed by Vasyliunas and Eviatar (2000) suggesting that O ions rather than H ions constitute this outflow, with similar densities as derived by Frank et al. (1997). These sets of observations of the thermal ionosphere were complemented by energetic ion measurements obtained by EPD/Galileo, showing that Ganymede is permanently embedded in an energetic plasma mainly composed of $S^+$ and $O^{n+}$ (Paranicas et al. 1999).

After Galileo flybys, a simple 1-D model of the atmosphere and ionosphere was proposed by Eviatar et al. (2001a) to fit PWS profiles with $O_2$ densities at the poles around $10^6$ $cm^{-3}$. The first attempt to reconstruct the origins and fate of Ganymede's neutral environment was described in Marconi (2007) who developed a 2-D axi-symmetric model of this atmosphere using a combination of Direct Simulated Monte Carlo and Test-particle Monte Carlo simulations focusing on the sunlit trailing hemisphere. He suggested neutral densities peaking around $10^9$ $cm^{-3}$ around the subsolar region and $10^8$ $cm^{-3}$ at the poles with the two main atmospheric species being $H_2O$ and $O_2$. Turc et al. (2014) developed the first 3-D model of Ganymede's atmosphere starting from Marconi (2007) simulation for the sunlit trailing hemisphere. These authors found essentially the same surface density but some discrepancies were found at high altitudes. Turc et al. (2014) found a Lyman α emission brightness up to 70 times smaller than observed by Barth et al. (1997), whereas Marconi (2007) had to multiply his simulated H Lyman α by a factor 4 to reproduce the observed intensities. Both models were able to reproduce the $O_2$ emission intensity observed by HST (Hall et al. 1998; Feldman et al. 2000). Furthermore, Plainaki et al. (2015) developed a 3-D Monte Carlo model of Ganymede's exosphere, including a model of ion precipitation in the auroral regions using a global MHD model of Ganymede's magnetosphere (Jia et al. 2009). They modeled Ganymede's sunlit leading hemisphere highlighting the different exospheric regions of the two main components $H_2O$ and $O_2$.

In this paper, we improve Turc et al. (2014) model, implementing a collisional scheme in order to identify the changes induced by atmospheric particle collisions. Marconi (2007) nicely illustrated the remarkable nature of this exosphere/atmosphere by plotting his estimate of the Knudsen number around Ganymede. Ganymede's atmosphere is predicted to be locally collisional (around the subsolar region) and collisionless elsewhere. Another aspect of Ganymede that was not described in Marconi (2007) or Turc et al. (2014), is how this atmosphere might change along Ganymede's orbit due to both its rotation and Jupiter's gravitational field. These new simulations take into account the revolution of Ganymede around Jupiter and Jupiter's gravitational field. In Section II, we present the main improvements of our model with respect to Turc et al. (2014). Section III presents how this work could contribute to a better understanding of Ganymede's atmospheric formation and evolution and is followed by a brief comparison with the few observations of this atmosphere and with other similar observations in our solar system in section VI.

**II Exospheric Global Model**

*II.1 Application to Ganymede*

A first version of the Exospheric Global Model (EGM) was applied to the description of Ganymede's atmosphere in Turc et al. (2014). This is a 3-D Monte Carlo model describing the fate of test-particles, ejected from a surface or an atmosphere of a planet or satellite, and subsequently followed under the effect of gravitational fields until their escape (when reaching an upper limit set at few planetary radii from the surface), their ionization/dissociation (by photo and/or electron impacts), or until their readsorption at the surface (which is species dependent). In order to reconstruct density, velocity, and temperature of the exospheric species (minor and major components), this model was parallelized **inducing** a very significant reduction of the restitution time and **the possibility to use** much larger total memory. These two improvements were crucial to describe the spatial and temporal distributions of all exospheric species in synchrony, that is, in the case of Ganymede's icy surface, all $H_2O$ products, namely $H_2O$, $H_2$, H, $O_2$, O and OH. The list of the reactions taken into account in our model is given in Table 2 of Turc et al. (2014) **and is reproduced in Table 1 below.**

| Reactions | Rate ($s^{-1}$) | Excess Energy (eV) | References |
|---|---|---|---|
| $H + hv \rightarrow H^+ + e$ | $4.5 \times 10^{-9}$ | 3.8 | Huebner et al. (1992) |
| $H_2 + hv \rightarrow H + H$ | $8.8 \times 10^{-9}$ | 2.2 | Marconi (2007) |
| $H_2 + hv \rightarrow H_2^+ + e$ | $3.1 \times 10^{-9}$ | 6.9 | Huebner et al. (1992) |
| $H_2 + hv \rightarrow H^+ + H + e$ | $6.9 \times 10^{-10}$ | 26 | Huebner et al. (1992) |
| $O + hv \rightarrow O^+ + e$ | $1.5 \times 10^{-8}$ | 24 | Huebner et al. (1992) |
| $OH + hv \rightarrow O + H$ | $9.7 \times 10^{-7}$ | 3.4 | Marconi (2007) |
| $OH + hv \rightarrow OH^+ + e$ | $1.6 \times 10^{-8}$ | 22 | Huebner et al. (1992) |
| $H_2O + hv \rightarrow H + OH$ | $5.2 \times 10^{-7}$ | 3.4 | Marconi (2007) |
| $H_2O + hv \rightarrow H_2 + O$ | $3.8 \times 10^{-8}$ | 3.4 | Marconi (2007) |
| $H_2O + hv \rightarrow H + H + O$ | $4.9 \times 10^{-8}$ | 4.6 | Marconi (2007) |

| Reactions | Rate (cm³ s⁻¹) | Excess Energy (eV) | References |
|---|---|---|---|
| $H_2O + h\nu \rightarrow H + OH^+ + e$ | $3.8 \times 10^{-9}$ | 21 | Huebner et al. (1992) |
| $H_2O + h\nu \rightarrow H_2 + O^+ + e$ | $5.2 \times 10^{-10}$ | 38 | Huebner et al. (1992) |
| $H_2O + h\nu \rightarrow OH + H^+ + e$ | $1.0 \times 10^{-9}$ | 28 | Huebner et al. (1992) |
| $H_2O + h\nu \rightarrow H_2O^+ + e$ | $2.1 \times 10^{-8}$ | 14 | Huebner et al. (1992) |
| $O_2 + h\nu \rightarrow O + O$ | $2.0 \times 10^{-7}$ | 1.3 | Marconi (2007) |
| $O_2 + h\nu \rightarrow O_2^+ + e$ | $3.0 \times 10^{-8}$ | 18 | Huebner et al. (1992) |
| $O_2 + h\nu \rightarrow O + O^+ + e$ | $8.4 \times 10^{-9}$ | 26 | Huebner et al. (1992) |
| **Reactions** | **Rate (cm³ s⁻¹)** | **Excess Energy (eV)** | **References** |
| $H + e \rightarrow H^+ + e + e$ | $9.1 \times 10^{-8}$ | 3.8 | Ip (1997) |
| $H_2 + e \rightarrow H + H + e$ | $9.6 \times 10^{-9}$ | 4.6 | Marconi (2007) |
| $H_2 + e \rightarrow H_2^+ + e$ | $1.6 \times 10^{-8}$ | 6.9 | Ip (1997) |
| $H_2 + e \rightarrow H^+ + H + e + e$ | $9.6 \times 10^{-10}$ | 26 | Ip (1997) |
| $O + e \rightarrow O^+ + e + e$ | $2.0 \times 10^{-8}$ | 24 | Ip (1997) |
| $OH + e \rightarrow O + H$ | $1.2 \times 10^{-8}$ | 3.4 | Ip (1997) |
| $OH + e \rightarrow OH^+ + e + e$ | $2.8 \times 10^{-8}$ | 22 | Ip (1997) |
| $H_2O + e \rightarrow OH + H + e$ | $3.7 \times 10^{-8}$ | 3.4 | Marconi (2007) |
| $H_2O + e \rightarrow H_2 + O + e$ | $1.6 \times 10^{-8}$ | 3.4 | Ip (1997) |
| $H_2O + e \rightarrow OH + H^+ + e + e$ | $4.3 \times 10^{-9}$ | 28 | Smyth and Marconi (2006) |
| $H_2O + e \rightarrow H + OH^+ + e + e$ | $4.0 \times 10^{-9}$ | 21 | Ip (1997) |
| $H_2O + e \rightarrow H_2 + O^+ + e + e$ | $7.1 \times 10^{-9}$ | 38 | Ip (1997) |
| $H_2O + e \rightarrow H_2O^+ + e + e$ | $2.1 \times 10^{-8}$ | 14 | Ip (1997) |
| $O_2 + e \rightarrow O + O + e$ | $1.3 \times 10^{-8}$ | 1.3 | Marconi (2007) |
| $O_2 + e \rightarrow O_2^+ + e + e$ | $2.0 \times 10^{-8}$ | 18 | Ip (1997) |
| $O_2 + e \rightarrow O + O^+ + e + e$ | $1.1 \times 10^{-8}$ | 26 | Smyth and Marconi (2006) |

**Table 1:** Photon and electron ionization and dissociation in Ganymede's atmosphere (Turc et al. 2014)

When impacting the surface, test-particles can either stick or be re-ejected at once. As in Turc et al. (2014) and Marconi (2007), electron ionization and dissociation impacts are described considering a uniform mono-energetic electron in the Alfven wings of Ganymede's magnetosphere (open field line regions, see section II.4) with an electron number density of 70 cm⁻³ and average energy of 20 eV. In the **closed** field lines region, we neglected ionization and dissociation by electron impacts. We assume that $O_2$ and $H_2$ species never stick to the surface, whereas all other species have a non-zero reabsorption probability.

Jupiter's gravitational force at the surface of Ganymede is equivalent to a few percent of Ganymede's gravitational force. However, because certain species have very long lifetimes in Ganymede's exosphere (e.g. $H_2$ and $O_2$), Jupiter's gravity may play a significant role in shaping its exosphere, as will be shown in Section III. Because we track molecules in Ganymede's reference frame, the Coriolis and centrifugal forces act on Ganymede's molecules in its atmosphere throughout the orbit. This is only relevant for certain species as the rotation period is ~7.2 days.

In Ganymede's inertial frame, the subsolar point rotates continuously, so that the day-to-night surface temperature gradient drives a global migration of the exospheric species towards the nightside. On the nightside, these particles can be readsorbed on the surface or thermally accommodate and be immediately re-emitted generating a peak in atmospheric density near the surface, as observed on the Moon (Hodges et al. 1974). We also introduce the possibility that returning volatiles can permeate the regolith or be transiently adsorbed. These can eventually return to the exosphere, a process potentially important for minor exospheric species, like Na, at Europa (Leblanc et al. 2005).

The standard outputs from this simulation are the 3-D values of the density, velocity, and temperature of all species considered. We therefore define a 3-D grid with 100 cells in altitude distributed exponentially from a cell of width 3 km in altitude at the surface to 300 km cell at 3 Ganymede radii ($R_G$), with 40 cells linearly spaced in longitude, and 20 cells in latitude, defined such that all cells have the same volume at a given altitude. We also reconstructed 2-D column density maps as derived from the 3-D neutral density, 2-D emission rates for the three main emission features observed around Ganymede (Lyman α and 130.4 and 135.6 nm oxygen) and a 2-D map of the surface reservoir. The model also calculates 3-D ionization rates by electron and photon impacts.

In the following, we describe the main improvements in the model of Turc et al. (2014), namely, a new thermal surface temperature model of the surface (Section II.2), a new description of sublimation (Section II.3), a new description of sputtering (Section II.4), the introduction of collisions (Section II.5) and a fusion scheme for the test-particles (Section II.6).

*II.2 Surface temperature*

Ganymede's surface temperature is estimated using a one-dimensional heat conduction model in which the thermal properties of the ice are treated as two layers (Spencer et al. 1989; available via

http://www.boulder.swri.edu/ spencer/thermprojrs). A low thermal inertia, $2.2\times10^4$ erg cm$^{-2}$ s$^{-1}$ K$^{-0.5}$, is assumed in the first few cm where the ice is heavily weathered, while the ice at depth has a much larger thermal inertia of $7\times10^4$ erg cm$^{-2}$ s$^{-1}$ K$^{-0.5}$, resulting in a longer thermal time scale. Density is set to 0.15 and 0.92 g/cm$^3$ in the top and bottom layers, respectively. The albedo is assumed to vary linearly over the body's longitude, from 0.43 on the leading hemisphere to 0.37 on the trailing (Spencer, 1987), and the emissivity is set at 0.96. A small correction for the latent heat of sublimation is included (Abramov and Spencer, 2008). Solar insolation over a synodic day is determined in degree increments of longitude with an eclipse length near equinox. This approach provides a basic description of longitudinal asymmetries about the subsolar axis that result from the thermal inertia of the ice, but does not account for any local variation of thermophysical parameters. Figure 1a shows the subsolar temperature as a function of orbital longitude, which results from shadowed sunlight during eclipse and albedo. Figure 1b maps the surface temperature at 270° orbital longitude (trailing) where a temperature of 146° K is the diurnal maximum.

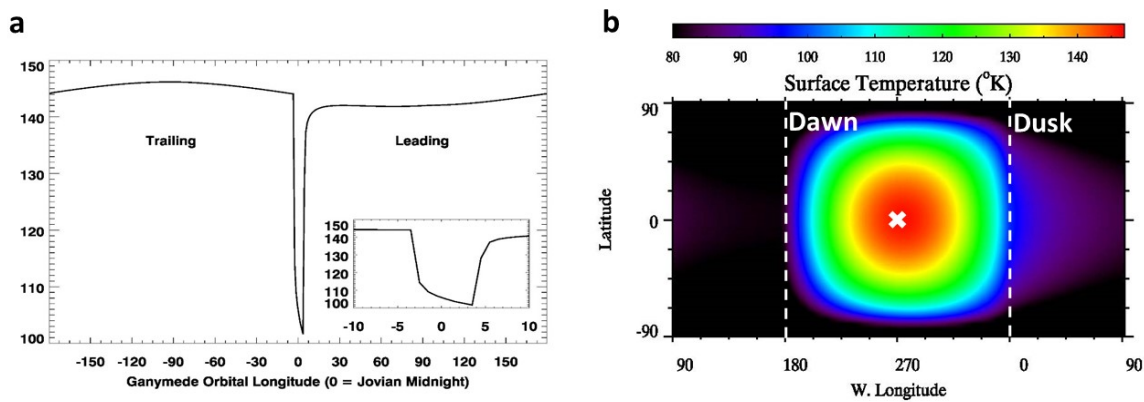

*Figure 1: a: Evolution of the subsolar surface temperature with respect to Ganymede orbital longitude (or phase angle). The insert in panel a is a zoom on the eclipse period showing the typical length of the eclipse and its impact on the subsolar surface temperature. b: Map of the surface temperature at 270° Ganymede phase angle with respect to latitude and West longitude. The subsolar point is at a latitude of 0° and a west longitude of 270° (white cross). Dawn and Dusk terminators are represented by the vertical dashed lines and labeled accordingly.*

As shown in Figure 1, the subsolar surface temperature peaks when Ganymede's trailing hemisphere is illuminated and is minimum at leading because of the change in surface albedo. The surface temperature displays a small but clear shift towards the afternoon due to lag from thermal inertia. The afternoon side (between 270° to 0° west longitude) of the surface is therefore significantly hotter than the morning side (between 180° and 270° west longitude).

*II.3 Sublimation*

In Marconi (2007), the sublimation rate at Ganymede's surface at a given temperature was given by

$$F = a \times T_s^{-0.5} \times \exp(-b/T_s) \qquad [cm^{-2} s^{-1}] \qquad (2)$$

with $a=1.1 \times 10^{31}$ $cm^{-2}$ $s^{-1}$ $K^{1/2}$ and $b = 5737$ K. Fitting the calculation by Johnson et al. (1981), we found different values for $a = 1.92 \times 10^{32}$ $cm^{-2}$ $s^{-1}$ $K^{1/2}$ and $b = 6146$ K. Fray and Schmitt (2009) published another formulation for this sublimation rate, also used by Vorburger et al. (2015). We fit the method of Vorburger et al. (2015) with equation (2) and derived values for $a=2.17 \times 10^{32}$ $cm^{-2}$ $s^{-1}$ $K^{1/2}$ and $b = 5950$ K (section II.2). These three sublimation rates are shown in Figure 2 below. We also considered a case where sublimation rate was significantly decreased (low sublimation case in Figure 2) to simulate an $H_2O$ exosphere dominated by sputtering. The Fray and Schmitt (2009) sublimation model represents an ideal case, based on experiments for pure crystalline ice. In practice, sublimation is highly dependent on the local grain structure and on an exact knowledge of the surface temperature, an error of 10 K giving a few orders of magnitude difference in sublimation rate. We therefore consider that even this low sublimation scenario might be within the realm of plausibility at Ganymede.

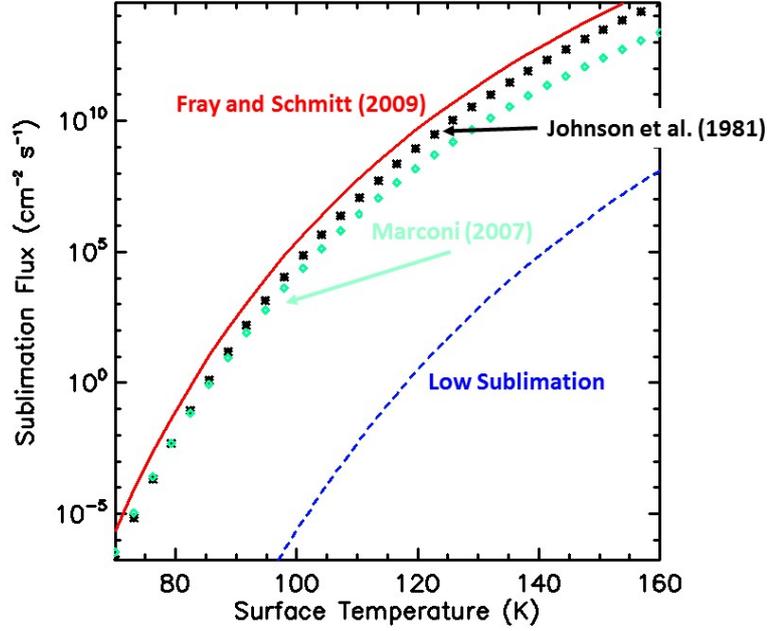

*Figure 2: Sublimation rate with respect to surface temperature as suggested by Fray and Schmitt (2009), red solid line, by Johnson et al. (1981), cross black symbols and by Marconi (2007) in square green symbols. The dashed blue line is the rate used for the low sublimation case throughout this paper.*

In the results presented below, we will use the Low sublimation case (Sections III, III.1 and III.2) to model the sublimation as well as Fray and Schmitt (2009) sublimation scenario (in Section III.3) to illustrate how a strong sublimation rate could change the structure and content of the exosphere.

*II.4 Sputtering*

Sputtering is due to energetic $S^{n+}$ and $O^{n+}$ Jovian particles that penetrate Ganymede's intrinsic magnetosphere through the polar regions and impact the polar surface regions (Cooper et al. 2001). This bombardment leads to the ejection of surface species, in particular $H_2O$, O, H, and OH, but also the production of $O_2$ and $H_2$ by radiolysis in the regolith in which the molecules leave with near thermal speeds. In order to describe the spatial distribution of this bombardment, in Turc et al. (2014), we arbitrarily limited the sputtering to latitudes above 45° as did Marconi (2007). However, based on the McGrath et al. (2013) auroral observations, we map out more precise regions where surface impacts might preferentially occur, using latitudinal limits of the auroral regions corresponding to the footprints of open field lines where Jovian plasma can penetrate through Ganymede's magnetosphere (Figure 3).

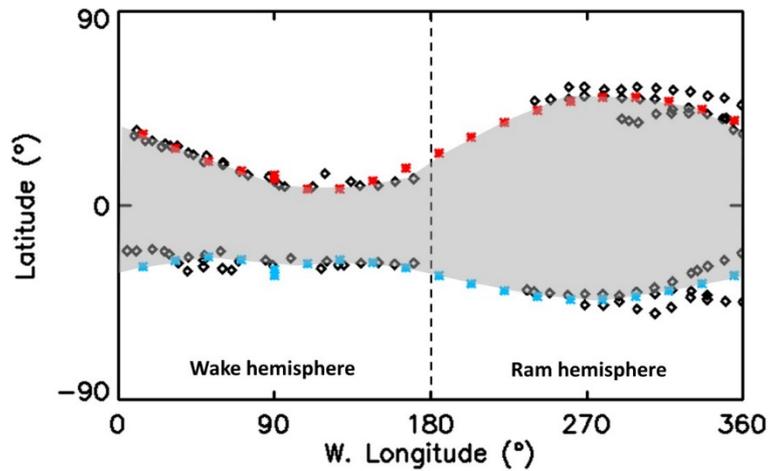

*Figure 3: Position of the peak auroral emission brightness as reported by McGrath et al. (2013) in a west longitude - latitude map (black symbols). The red and blue symbols represent the polynomial fit performed in our model to reproduce the closed/open field line boundaries in each hemisphere. The magnetosphere is most compressed in the ram direction at a longitude of 270°, corresponding to a higher latitude boundary. The grey region corresponds closed Ganymede field lines where we assume negligible sputtering.*

By excluding the closed field line regions, we neglect any surface production of exospheric particles by sputtering here, because according to Fatemi et al. (2016) and Leclercq et al. (2016), the total precipitation rate in the closed field line region is much smaller than the total precipitation rate in the opened field line region (see also Plainaki et al. 2015). However, ions reaching the closed field line region are more energetic and should, on average, correspond to particles that sputter with higher yield, an issue that has not been addressed in Fatemi et al. (2016) and Leclercq et al. (2016). In the following, we neglect this contribution for simplicity. In the open field line regions, we assume a uniform incident ion flux on the surface as suggested by Fatemi et al. (2016) and Leclercq et al. (2016). No variation of the precipitating flux with respect to Jupiter plasma centrifugal equator or sheet has been included in this modelling for simplicity.

To calculate the efficiency of an impacting ion to sputter surface particles, we include the surface temperature dependence of the yield by Cassidy et al. (2013). Following their approach, we used Johnson et al. (2009) and Fama et al. (2008) approximations of the yield (number of molecules and atoms ejected per impacting particle) with respect to the incident energy:

$$Y_{H2O}(T) = Y_{H2O}(T_0) \times (1 + 220 \times \exp(-0.06 \text{ eV}/k_B T)$$

(3)

and

$$Y_{O2}(T) = Y_{O2}(T_0) \times (1 + 1000 \times \exp(-0.06 \text{ eV}/k_B T)$$

(4)

for sputtered $H_2O$ molecules (equation (3)) and $O_2$ molecules (equation (4)). From these two equations, and knowing the expected flux of ion impacting the surface, we calculate the sputtered flux from Ganymede's icy surface. However, the value of $Y_{O2}(T_0)$ for $O_2$ and $Y_{H2O}(T_0)$ for $H_2O$ needs to be estimated as it depends on many parameters, such as the energy and charge of the impacting particle, the surface structure, and the non-ice components (Teolis et al. 2016). Another approach is to choose $Y(T_0)$ in order to reproduce the observed quantities in Ganymede's exosphere. For instance, the observed FUV $O_2$ auroral emission can be directly used to constrain $Y(T_0)$ (see also the discussion in Section IV). Based on such observations, we can infer an effective yield of $O_2$ molecules and then derive corresponding effective yields for the other water products H, O, OH, $H_2$, and $H_2O$ from their known relative yields with respect to $O_2$.

Marconi (2007) considered relative yield values equal to 1 for $H_2O$, 1/18 for H, O, OH, to 16 for $H_2$ and to 8 for $O_2$, whereas Smyth and Marconi (2006) suggested values equal to 1 for $H_2O$, 1/17 for H, O, OH, to 2 for $H_2$ and to 1 for $O_2$ at Europa. At Europa, Shematovich et al. (2005) considered that 1 molecule of $H_2O$ should be ejected for each 1/10 $O_2$ molecules. Noting the wide range in previous models, we choose to base our yields on Cassidy et al. (2013), who made the first detailed calculation of sputtering rates at Europa based on laboratory experiments, Europa's expected surface temperature, and the expected intensity and spatial distribution of the impacting ions flux. These authors estimated that $2\times10^{27}$ $H_2O$/s and $10^{26}$ $O_2$/s should be ejected from Europa's surface. In a companion paper, we applied the EGM model to Europa's atmosphere (Oza et al. 2016) and thereby constrained $Y(T_0)$ for $O_2$ and $H_2O$ from the Cassidy et al. (2013) calculations of the total rates ejected from Europa's surface. According to Oza et al. (2016), each time a molecule of $H_2O$ is ejected from the surface, an average of 1/20 $O_2$ is ejected from Europa's surface. The relative yield for $H_2$ should be nearly stoichiometric, i.e. twice that of $O_2$, since they dominate the product composition and are both formed by radiolysis of the icy surface. H, O, and OH yields should be small with respect to $O_2$, $H_2$ and $H_2O$ yield and are set to $1/20^{th}$ of the $O_2$ yield in agreement with Smyth and Marconi (2006). These yields represent a weighted average over all charge states of O and S ions sourced in Io's plasma torus, which bears little compositional differences between 9.4 and 15 Jovian radii (Delamere et al., 2005).

When an exospheric particle is ejected from the surface by sputtering, we proceed in the same way as in Turc et al. (2014). Namely, $H_2O$, O, H, and OH are ejected from the surface following an energy distribution described in equation (8) in Turc et al. (2014) and following a Maxwell-Boltzmann distribution at the local surface temperature in the case of $O_2$ and $H_2$ (equation 7 in Turc et al. 2014). In order to properly describe the entire energy distribution up to the escape energy, we bin these two energy distributions into 15-to-25 energy intervals (from 0 to an energy slightly above the escape energy) and simultaneously eject 15 to 25 individual test-particles with a weight calculated as the integral of the energy distribution over the energy range of each bin. Teolis et al. (2016) suggested that $O_2$ molecules could be ejected following the sum of two distributions: 70% of the $O_2$ being ejected following a Maxwellian Boltzmann distribution and 30% being ejected following a distribution in $E/(U+E)^3$ (Johnson et al. 1983). Such a choice would populate the higher altitudes slightly more efficiently than in our approach but would not change the main conclusions of this paper.

*II.5 Collisional scheme*

Because in some regions of Ganymede's atmosphere, collisions between atmospheric particles might shape the density (Marconi 2007), we also include collisions between molecules and atoms. At each time step, the total density of each species in a cell is calculated. From these densities, the maximum number of collision pairs between each test-particle and the other test-particles for a given species in the cell is calculated using the approach described in Bird (1994) with cross sections taken from Lewkow and Kharchenko (2014). These authors developed a new scheme including cross sections depending on both energy and angle that allows an accurate description of collision between any species colliding with energies below a few tens of eV. Typical collision cross sections for molecule-molecule or molecule-atom collision at 1 eV are $4.8 \times 10^{-15}$ cm$^{-2}$ and do not depend **neither** significantly on energy between 0 to few eVs **nor on the species considered in this work (H, H$_2$, O, OH, O$_2$ and H$_2$O).** These cross sections are close to the ones used by Marconi (2007), as an example, equal to ~$6.9 \times 10^{-15}$ cm$^{-2}$ for H$_2$O-H$_2$O collision or to ~$3.9 \times 10^{-15}$ cm$^{-2}$ for O$_2$-O$_2$ collision. We then randomly select a number of test-particle pairs equal to the expected maximum number of collisions that should occur according to Bird (1994). One of the difficulties when including minor species is that test-particles, in a given cell, might have very different weights (i.e. they might represent vastly different number of real particles). We therefore developed a new scheme to treat such collisions. For a test-particle A of weight $W_A$ and species $S_A$, we randomly determine with which species, $S_B$, a test-particle A will most likely collide. We then randomly select $n_A$ test-particles $B_i$ of weight $W_{Bi}$ of the species $S_B$ in order to have $\sum_{i=1}^{nA} W_{Bi} \geq W_A$. For each pair of test-particles $B_i$ and A, we then evaluate the likehood of a collision taking into account the relative velocity between A and $B_i$. If a collision occurs, we determine new velocity vector after the collision, V'$_{Ai}$ and V'$_{Bi}$ following Lewkov and Kharchenko (2014). We then create a new test-particle $A_i$ with velocity V'$_{Ai}$ and weight $W_i$ and repeat that scheme $n_A$ times. The weight of the particle A is then reduced by $W_{Bi}$. The velocity of each $B_i$ test-particle after collision is equal to V'$_{Bi}$ with the exception of the $n_A^{th}$ $B_i$ test-particle which is divided into two particles: one with velocity V'$_{BnA}$ and weight W'$_{BnA}$ = $W_A - \sum_{i=1}^{nA-1} W_{Bi}$ and the other with velocity V$_{BnA}$ and weight W$_{BnA}$ - W'$_{BnA}$.

All collisions occur with energies below 10 eV so that molecular dissociation is negligible.

*II.6 Fusion scheme*

Since the collisional scheme implies the creation of new particles (see section II.5), we also had to implement a fusion scheme in order to limit the number of test-particles followed by the model at a given time. We tested two approaches to merge test-particles, both suggested by Lapenta (2002), the binary scheme, and the tertiary scheme. The binary scheme is faster and since it leads to similar results, we employ it. The binary scheme consists in selecting a pair of test-particles A and B which are close each other in phase space (with positions $q_A$ and $q_B$, velocities $V_A$, $V_B$ and weight $W_A$, $W_B$ respectively) of the same species and in the same cell that minimizes $(W_A + W_B)(V_A^2 - V_B^2)$. We then create a new particle with weight equal to $W=W_A+W_B$, position equal to $q = (W_A q_A + W_B q_B)/(W_A+W_B)$ and velocity equal to $V = (W_A V_A + W_B V_B) / W$.

For every arbitrarily fixed number of time steps (typically one hundred time steps), we count the number of particles of each species within one cell. If this number is larger than a threshold, defined as to avoid too many test-particles of a given species in a cell, the fusion scheme is activated, adapting to the computational load this presents. Test-particles are therefore selected following the approach described above, and are merged as long as this number continues to exceed the threshold.

**III Ganymede's atmosphere**

In the following, we will use the phase angle of Ganymede (equivalent to its sub-observer longitude) to indicate its position around Jupiter. A phase angle of 0° corresponds to Ganymede being in the shadow of Jupiter at midnight local time (eclipse), whereas phase angles of 90° and 270° correspond to the sunlit leading hemisphere apex and sunlit trailing hemisphere apex respectively. Positions at the surface are defined using west longitude (increasing with local time), 0° planetary longitude being the subsolar point at a 0° phase angle. The simulation is performed in a GphiO rotating frame with the x-axis always pointing along the corotational direction, y-axis pointing towards Jupiter, and the z-axis pointing towards Ganymede's north pole (the x and y axis are inside the orbital plane of Ganymede). Section III.1 is dedicated to the description of the fate of Ganymede's exosphere along its rotation around Jupiter, without taking into account the effect of collision between atmospheric particles, an issue specifically discussed in Section III.2. Section III.3 is based on simulations where collisions were also neglected.

Using the sputtering description in section II.4, the typical surface temperatures of Figure 1, and the low sublimation rate as described in Figure 2, the global sublimated flux is equal to $8.0 \times 10^{21}$ $H_2O$/s whereas the average ejected flux by sputtering is equal to $8.0 \times 10^{27}$ $H_2O$/s, $3.6 \times 10^{26}$ $O_2$/s, $7.2 \times 10^{26}$ $H_2$/s and $1.0 \times 10^{25}$ H, O and OH/s. Marconi (2007) and Turc et al. (2014) used a flux of $1.5 \times 10^{26}$ $H_2O$/s and $1.2 \times 10^{27}$ $O_2$/s for sputtering, whereas Plainaki et al. (2015) used a flux of $7 \times 10^{25}$ $H_2O$/s for sputtering and $7 \times 10^{29}$ $H_2O$/s for sublimation using the same model for sublimation as in Marconi (2007) and Turc et al. (2014) (green square symbols in Figure 2).

We simulated more than 4.5 rotations of Ganymede around Jupiter in two weeks on 64 CPUs (that is, $3.7 \times 10^6$ time steps of 0.75 s each for a total of 780 simulated hours). During that run, $2.6 \times 10^7$ $H_2O$ test-particles (simulating a total of $2.2 \times 10^{34}$ $H_2O$ molecules) and $2.1 \times 10^7$ $O_2$ test-particles (for $10^{34}$ $O_2$ molecules) were ejected from the surface. Of these $1.2 \times 10^6$ $H_2O$ test-particles escaped the simulation domain, $2.4 \times 10^7$ were reabsorbed in the surface, $6 \times 10^5$ were dissociated or ionized and none were merged (meaning that the maximum number of test-particles in a cell allowed by the simulation was never reached for $H_2O$). For $O_2$, $1.3 \times 10^6$ were dissociated or ionized, $1.9 \times 10^7$ were merged. The $3 \times 10^5$ $O_2$ test-particles which escaped the simulation domain were generated from the tail of the energy distribution ejected from the surface. Being associated with very small weight, these escaping molecules represented an escaping flux of $6 \times 10^{17}$ $O_2$/s. Merging test-particles is much more important for $O_2$ than for $H_2O$, because $O_2$ test-particles are essentially confined near the surface. At each time, a total number of 1 to $2 \times 10^6$ test-particles are followed. After convergence towards a steady state, the typical escape rates are $1.6 \times 10^{27}$ $H_2O$/s, $6 \times 10^{17}$ $O_2$/s, $1.7 \times 10^{26}$ O/s, $1.2 \times 10^{26}$ OH/s, $3.0 \times 10^{26}$ $H_2$/s and $2.3 \times 10^{26}$ H/s.

*III.1 Seasonal variation of Ganymede's exosphere*

A main improvement with respect to previously published models of Ganymede's atmosphere is that we describe the evolution of Ganymede's exosphere throughout Ganymede's orbit around Jupiter while taking into account the influence of Jupiter's gravitational field. By simulating a period of time long enough to cover several Ganymede orbits, our model simultaneously calculates the motions of all individual test-particles in Ganymede's inertial frame as well as the motion of Ganymede around Jupiter. In general, only three orbits are needed to achieve a steady state exosphere on Ganymede. Steady state is evaluated at a fixed phase between two successive orbits for a given time t, when the reconstructed density, velocity, and temperature of the $O_2$ and $H_2O$ exospheric components and their loss rates vary by less than a few percent. In the results presented in this section, collisions are not taken into account for the following results and will be specifically discussed in section III.2.

Most previous models have focused on one specific position of Ganymede corresponding to fixed solar illumination and ram directions: the trailing illuminated hemisphere case in Turc et al. (2014) and Marconi (2007), and the leading illuminated hemisphere case in Plainaki et al. (2015). However, Ganymede's rotation implies that the geographical regions where sputtering and sublimation rates reach a maximum, rotate with respect to one another. One direct consequence of this rotation is a variable sputtering efficiency since yields depend strongly on the surface temperature and, therefore, on the solar zenith angle (Cassidy et al. 2013).

The typical time for $O_2$, ejected at the equator to migrate from one hemisphere to the other (that is to move from a position at the surface to its opposite position on the other side of the satellite) can be roughly estimated following Hunten et al. (1988). According to these authors, the collisionless migration time of a particle moving in a surface-bounded exosphere is on the order of $t_m = t_b \lambda^2$ where $t_b = 2(H/g)^{1/2}$ is the ballistic hop time and $\lambda = R_G/H$. $R_G$ is Ganymede's radius and H is the scale height $H = k_B T_S / (mg)$ with $k_B$ the Boltzmann constant, $T_S$ the surface temperature (as defined in section II.2), m the mass of the particle and g the gravitational acceleration at the surface. For an $O_2$ ejected from noon at the sunlit trailing hemisphere to migrate across the hemisphere is, $t_m \sim 765$ hours or more than 4 orbital periods using Ts=143.5 K (see Figure 1a), H = 26 km, $\lambda$=101, $t_b$=270 s. Migration across the hemisphere takes $t_m \sim 1690$ hours, or more than 9 orbits using 85K, H=15 km, $\lambda$=171, $t_b$=208 s. Precise migration times are not very meaningful, since $t_m$ well exceeds the diurnal timescale of the thermal wave, yet, the calculation is useful to conceptualize the atmosphere's diffusion. Given that the typical lifetime of an $O_2$ is on the order of one orbital period, newly ejected $O_2$ roughly fixed in Ganymede's reference frame moves therefore in local time due to Ganymede's rotation. The typical time scale for the formation of the $O_2$ exosphere can be estimated by dividing the average vertical column density ($2.4\times10^{14}$ $O_2/cm^2$) by the average ejection rate ($3.5\times10^8$ $O_2/cm^2/s$) or ~190 hours, that is, more than one orbital period. Therefore, the $O_2$ exosphere at a given orbital position will depend on its history and evolution over about one Ganymede revolution. Because the $H_2O$ molecules are quickly reabsorbed in the surface, the timescale of the $H_2O$ exosphere is much shorter than one rotational period. In the case of $H_2$, due to its low mass, the $H_2$ exospheric will rapidly form across Ganymede becoming spatially uniform on timescales much shorter than an orbital period and is, therefore, less dependent on localized production.

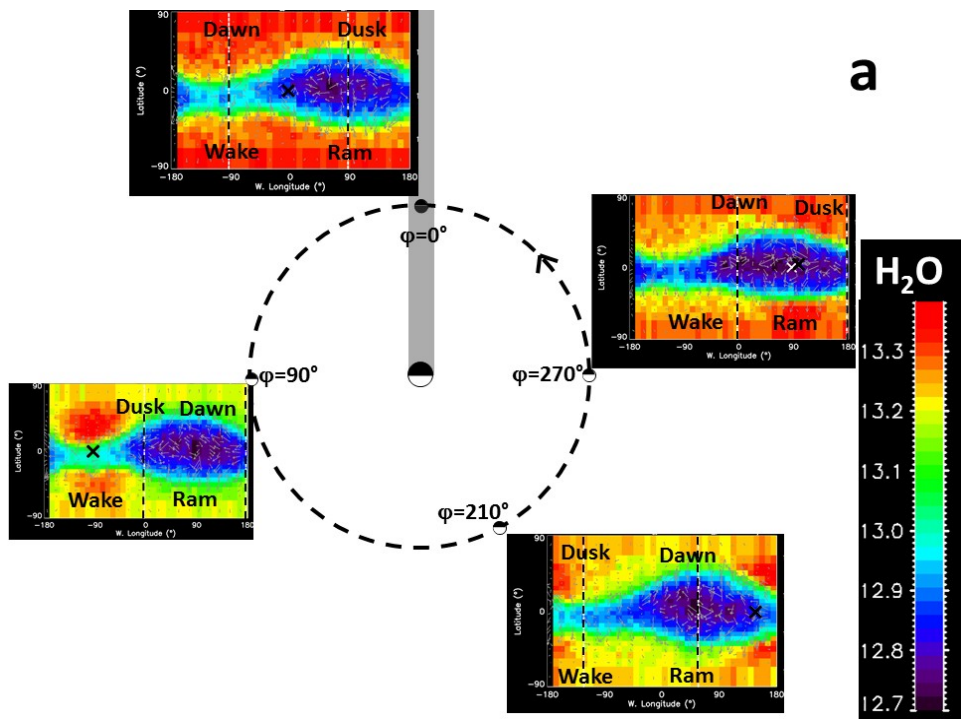

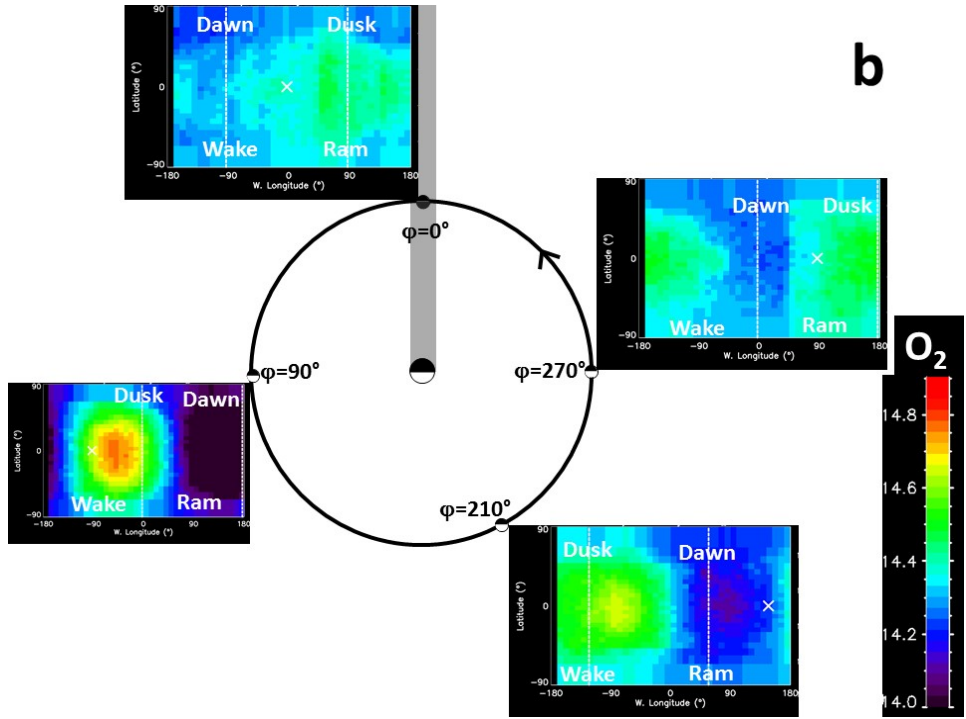

*Figure 4: Vertical column densities ($\log_{10}$ cm$^{-2}$) of the H$_2$O (panel a) and O$_2$ exospheres (panel b) at different Ganymede phase angles (90°, 210°, 270° and 0°) with respect to latitude (vertical axis) and west longitude (horizontal axis) in the low-sublimation scenario. The subsolar point is indicated by the black or white cross on each panel, the dusk and dawn terminators by vertical dashed black or white lines. The ram and wake hemispheric centers are also indicated. Ganymede's direction of rotation is indicated by the counter-clockwise arrow in the plane of Jovian co-rotation. Jupiter is at the center of each figure.*

In Figure 4, we display the vertical column densities of H$_2$O (panel a) and O$_2$ (panel b) at four positions around Ganymede. Each figure corresponds to the average simulated density over a 6° interval centered on the indicated phase angle value. This interval allows the simulation to reach a reasonable signal-to-noise ratio for the reconstructed densities. The H$_2$O vertical column density peaks most of the time (except at 270° phase angle) on the plasma-wake hemisphere (also the leading hemisphere) because H$_2$O is predominantly produced by sputtering. In the low sublimation rate scenario displayed here, sputtering should occur at lower latitudes on the wake hemisphere than on the ram hemisphere (as suggested by the opened/closed field lines boundary in Figure 3), more H$_2$O is therefore ejected on the wake hemisphere than on the ram hemisphere. This magnetospheric effect, when convolved with the strong thermal dependence of radiolysis (Equation 3) and sublimation (Figure 2), effectively increases the $H_2O$ ejection rate at the sunlit leading hemisphere/plasma wake (see also Plainaki et al. 2015). We therefore simulate a slightly denser $H_2O$ exosphere on the leading illuminated hemisphere position (radial column density of $1.5 \times 10^{13}$ $H_2O/cm^2$) than on the trailing illuminated hemisphere position (radial column density of $1.3 \times 10^{13}$ $H_2O/cm^2$). Observed from the Earth, the average line-of-sight (LOS) column density at a phase angle of 90° over the hemisphere disk would be equal to $3.4 \times 10^{13}$ $H_2O/cm^2$, whereas it would decrease to $2.9 \times 10^{13}$ $H_2O/cm^2$ at a phase angle of 270° (see also Figure 6). $H_2O$ in the equatorial plane has densities smaller than $10^5$ $H_2O/cm^3$ coming essentially from the polar region, whereas densities at the poles, are greater by two orders of magnitude. Since $H_2O$ molecules efficiently stick to the surface, the polar $H_2O$ molecules do not migrate efficiently towards the equatorial regions and remain confined around their ejection region.

The $O_2$ exosphere is significantly different at the equatorial regions because the probability for $O_2$ to stick to the surface, even at coldest nightside surface temperatures, is low. Consequently, the $O_2$ exosphere is spread across all longitudes and latitudes. There is however a clear dusk/dawn asymmetry of the $O_2$ spatial distribution with a peak in density at afternoon between $\varphi=90°$ to $\varphi=200°$, Ganymede local time and in the early night between $\varphi=90°$ to $\varphi=210°$. The average radial column density varies from $2.2 \times 10^{14}$ $O_2/cm^2$ at a phase angle of 20° to a peak of $2.5 \times 10^{14}$ $O_2/cm^2$ at a phase angle of 150°. Both the asymmetry between dusk/dawn terminators and the variation of the exospheric content are a clear result of the production and loss timescales of the $O_2$ exosphere along Ganymede's orbit and can be understood by the variation of the ejection pattern of the $O_2$ particles as described below and in Figure 5.

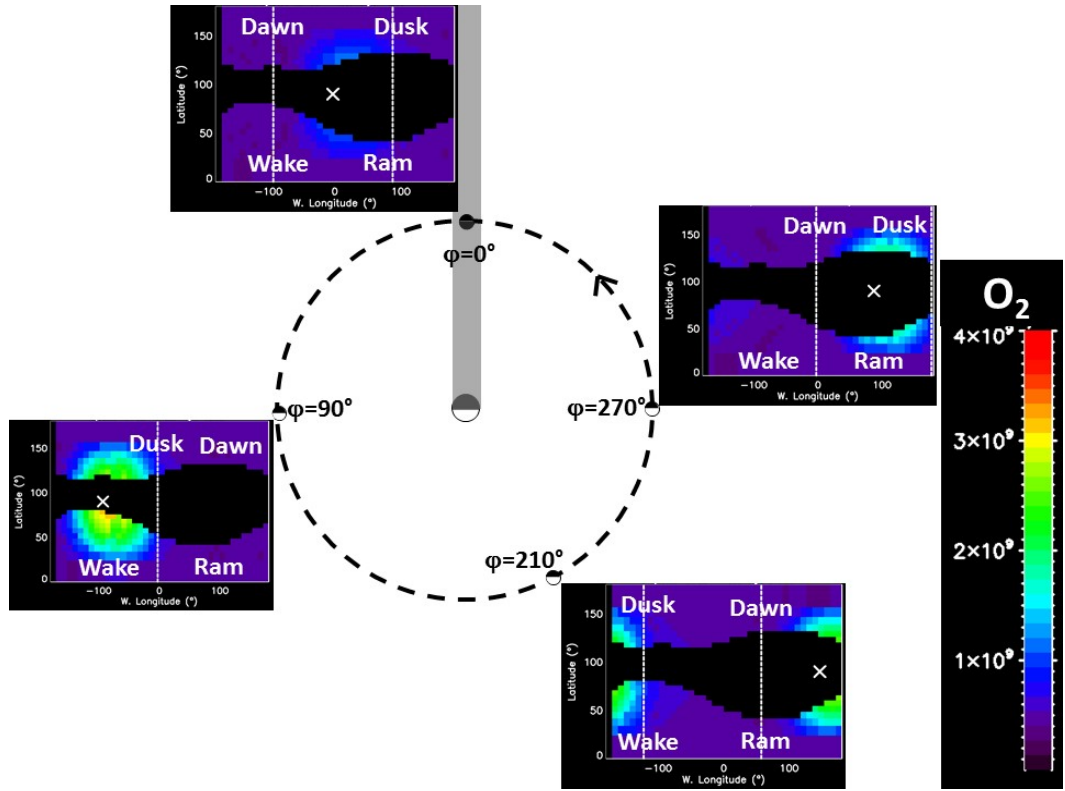

*Figure 5: West longitude - latitude maps of the rate of $O_2$ ejection (in $O_2/cm^2/s$) by sputtering at various Ganymede phase angles. On each figure the X indicates the subsolar point whereas the vertical dashed lines represent the terminators labeled accordingly as dawn and dusk. Also indicated on each panel are the positions of the wake and ram hemispheres.*

Due to the thermal inertia of the surface, the role of the ram direction for the sputtering and the shape of the opened field lines region, there is a larger flux ejected close to the dusk than to the dawn when averaging over an orbit. The peak $O_2$ source rate occurs around a phase angle of 90° (at $5\times10^8$ $O_2/cm^2/s$) due to the variation in latitude of the opened/closed field lines boundary from the ram to wake hemispheres. At a phase angle of 90° the largest open field line region (wake hemisphere) is also where the surface temperature is highest resulting in an enhanced sputtering efficiency according to equations (3) and (4). In contrast, at a phase angle of 270°, the opened field line region is smallest on the dayside (the ram hemisphere), leading to a globally lower rate of ejection ($3\times10^8$ $O_2/cm^2/s$). Sputtering ejection reaches a minimum at a phase angle of 0° ($2.4\times10^8$ $O_2/cm^2/s$) because of the eclipse induced by Jupiter's shadow and the rapid decrease of Ganymede's surface temperature (Figure 1a). Since the modeled atmosphere is nearly stationary, a variation in the ejection or loss rate produces a variation in the total content of the $O_2$ exosphere. Therefore, the total $O_2$ content reaches a minimum around a phase angle of 20° due to the low ejection rate during the previous half of the rotation, between phase angles 200° to 20° with a peak around a phase angle of 150° following the maximum in ejection rate around a phase angle of 90°. That is, Ganymede's $O_2$ exospheric total content decreases during a portion of its orbit centered around a phase angle of 330° where the loss rate due to electron impact ionization/dissociation is larger than the ejection rate. During this portion of the orbit, Ganymede's decaying exosphere is a residual that formed earlier and conversely, during the following portion of the orbit around a phase angle of 150°, the total content increases. The $O_2$ exosphere also appears slightly more extended on the dayside (Figure 4) due to larger surface temperatures. In our model, $O_2$ is mainly ejected from the higher latitudes by sputtering as explained in section II.4 (Figure 5). However, the vertical column density of the $O_2$ clearly peaks at the equator (Figure 4b) and not at high latitudes. The $O_2$ diffusion time suggests that the molecules remain close to their source regions. However, this does not take into account the effect of the centrifugal force which tends to drive molecules equatorward. This force has an average intensity equal to few percent of Ganymede's gravitational force at the surface and can therefore, shape the $O_2$ exosphere as seen in Figure 6a. The observed dusk/dawn asymmetry seen in Figure 4a leads also to larger column density on the right (dusk) side of the exosphere as seen from the Sun in Figure 6a. This dusk/dawn ratio in the $O_2$ column density peaks at a value of 3.4 at a phase angle of 150°. Ganymede's exosphere near the leading hemisphere portion of the orbit is formed by the production and spreading of the high latitude ejecta as seen from Figure 5. The dusk/dawn asymmetries simulated here are also produced when simulating the orbital evolution of Europa's $O_2$ exosphere using the EGM (Oza et al. 2016), in accordance with recent HST orbital observations of auroral features there (Roth et al. 2016)

The H exosphere is produced from the polar region by sputtering in our model. The column density, as seen from the Sun shown in Figure 6b, reflects the source region and that the ejection energy primarily leads to escape. Sputtered $H_2$, on the other hand, is ejected at the local surface temperature at lower energies, and with its higher mass a significant fraction remains gravitationally bound. Since it does not stick, a nearly isotropic $H_2$ exosphere results as shown in Figure 6c (despite electron and photon impact dissociation of $H_2$). In the low-sublimation scenario presented in Figure 6, $H_2O$ is primarily ejected from polar regions with energies comparable to escape energy. Therefore $H_2O$ tends to populate high altitudes with column densities peaking near the regions of sputtered production as indicated in Figure 6d.

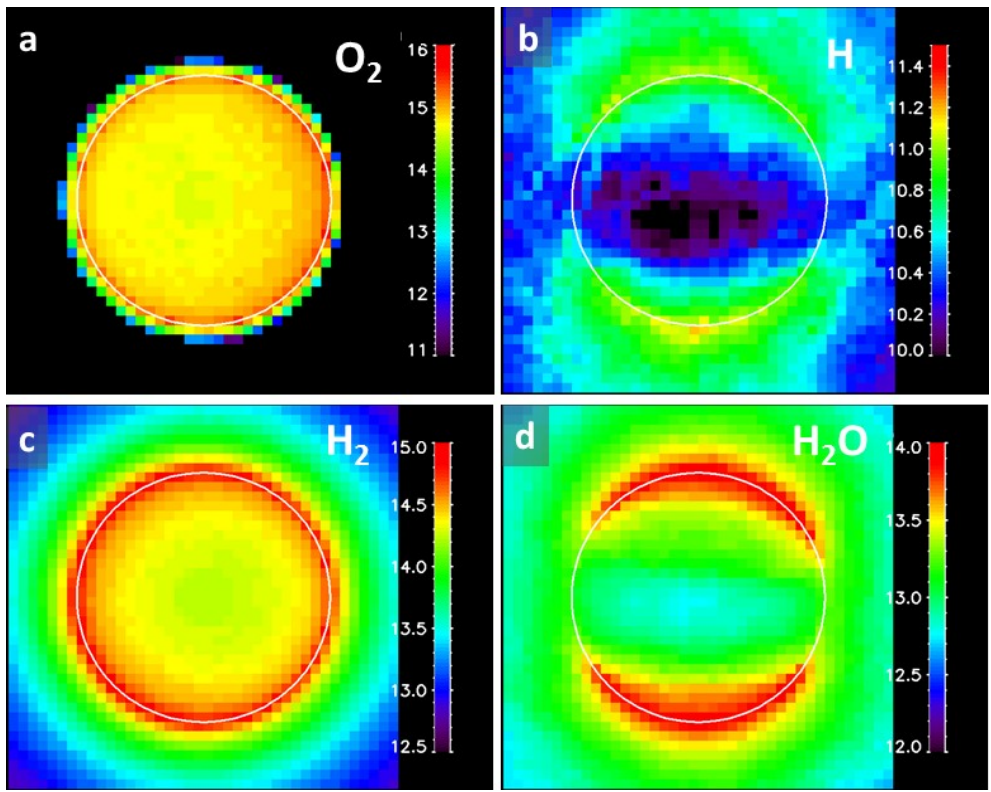

*Figure 6: Reconstructed column densities (in $\log_{10}$ $cm^{-2}$) of the $O_2$ exosphere (panel a), of the H exosphere (panel b), $H_2$ (panel c) and $H_2O$ (panel d) at a phase angle of 270° as seen from the Sun.*

We also tested if the formation of a surface reservoir by the readsorped $H_2O$ test-particles subsequently released again could induce a change of the exosphere density. However, because $H_2O$ is the main component of the surface, we did not notice any effect associated with the formation of an exospheric reservoir of $H_2O$ in the surface.

*III.2 The role of collisions*

As suggested by Marconi (2007), Ganymede's atmosphere might be locally collisional. We therefore investigated the effect of momentum transfer collisions on the global distribution of the exosphere. Since the implementation of collisions, requires of the order of a few weeks on 128 CPUs, we only ran simulations on one phase angle. Starting from the steady-state solution calculated in the collisionless regime (as described in section III.1), we simulated the fate of Ganymede's exosphere, including collisions (section II.5), during a time period long enough (few tens of hours) to reach an accurate description of the exosphere up to ~1200 km in altitude. In Figure 7, we compare the density profile of the two main components in Ganymede's exosphere with and without taking into account the effects of collisions at a phase angle of 90°.

As shown in Figure 7, the main effect of collisions on the density profile is to enhance the population of $O_2$ at high altitudes. Indeed, most of the collisions occur either between $O_2$ molecules or between $O_2$ and $H_2O$ molecules, the two main species in the collisional region near the surface (within the first tens kilometers). In this layer, collisions between other thermal $O_2$ molecules have a negligible effect on the $O_2$ radial density distribution. However, $O_2$ collisions with the more energetically sputtered $H_2O$ possessing a high energy tail (~$1/E^2$) can add significant energy to the $O_2$ molecules, thereby expanding the $O_2$ exosphere.

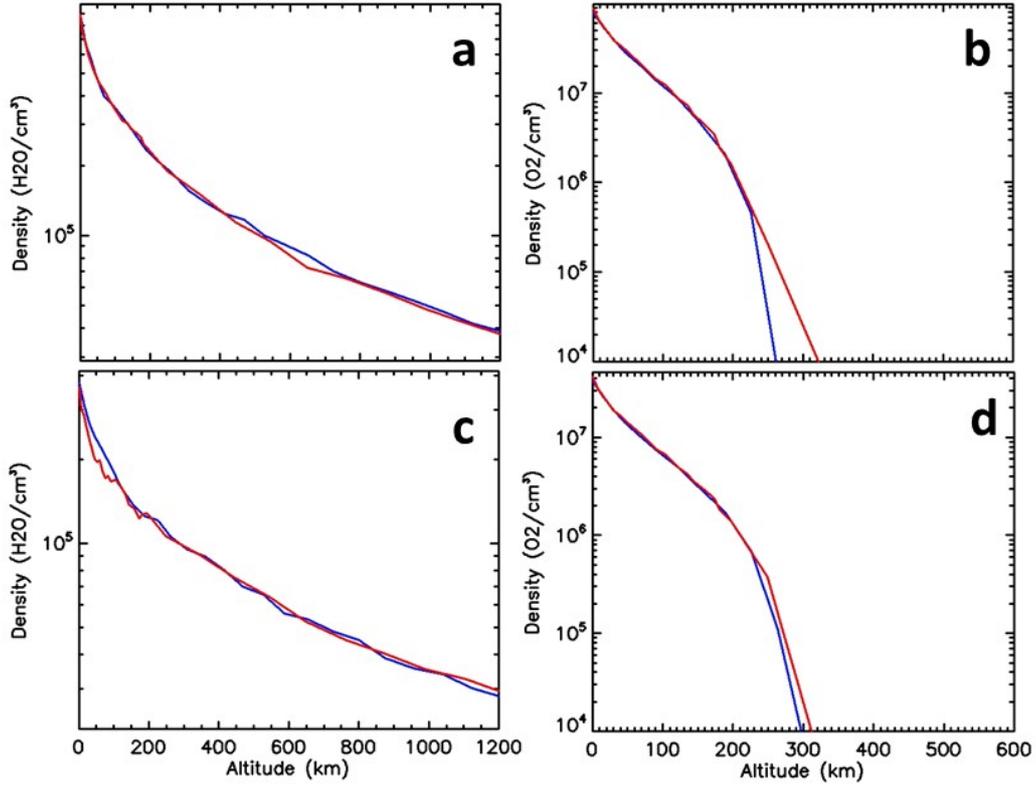

*Figure 7: Density profiles simulated by EGM without collision (blue lines) and with collision (red lines) as calculated at a phase angle of 90°. Panels a and b are for $H_2O$ and $O_2$ exospheric species respectively at the subsolar point whereas panels c and d are for $H_2O$ and $O_2$ exospheric species respectively at the polar region. The density profile is calculated as the average density within 10° of the subsolar point (panels a and b) and to the North pole (panels c and d).*

We find that our results are in good agreement with the conclusions by Shematovich (2016) model of Ganymede's exosphere. This author found that collisions caused the 10 to 100 km altitude region to be populated by both $O_2$ and $H_2O$. Similarly, Marconi (2007) suggested that charge exchange between $O_2^+$ ion and $O_2$ molecules could enhance the population of $O_2$ in the upper exosphere. Charge exchange was also investigated by Dols et al. (2016) and by Luchetti et al. (2016) at Europa who both suggested that charge exchange between $O_2^+$ and $O_2$ might be the dominant loss process for that satellite. Including charge exchange accurately requires coupling between a magnetospheric and an exospheric model, which is work in progress. The near surface density has a typical scale heights for $O_2$ of 50 km which is significantly larger than the expected scale height, 27 km, for an atmosphere in thermal equilibrium with the surface at a maximum temperature of 146 K. In this model, $O_2$ molecules are always ejected from the surface following a Maxwell-Boltzmann distribution defined by the surface temperature. However, as discussed in section III.1, $O_2$ molecules are particularly sensitive to the non-inertial forces because of their long residence time in the atmosphere. A large portion of the $O_2$ molecules come from regions far from the local surface leading to a globally uniform layer at the poles. This was illustrated in Turc et al. (2014) where the $O_2$ density at the poles decreased much faster (with scale height around 22 km) than in the subsolar region.

*III.3 The role of sublimation*

For the Fray and Schmitt (2009) parametrization of sublimation, the total ejection rate would be equal to $3.5 \times 10^{29}$ $H_2O$/s (comparable to the $8.0 \times 10^{27}$ $H_2O$/s sputtered and to the $7 \times 10^{29}$ $H_2O$/s sublimated in Plainaki et al. (2015)). This implies that the $H_2O$ exosphere would be dominated by sublimation. This is clearly illustrated by simulation results in Figure 8a which displays the exospheric density in the equatorial plane, with the same format as in Figure 4a. The $H_2O$ density peaks as expected around the subsolar region with a clear shift towards the afternoon due to the shift of the maximum temperature on the dayside (Figure 1b). The eclipse period is particularly dramatic, since as shown by Turc et al. (2014), the quick decrease of the surface temperature (Figure 1a insert) leads to an almost complete disappearance of the $H_2O$ exosphere (characterized by a collapse of the density by more than 4 orders of magnitude). The extended part of the $H_2O$ exosphere (in longitude and altitude with density larger than $10^4$ $H_2O$/cm$^3$) is due to the sputtering of polar regions by the incident Jovian plasma. It is clearly more important on the wake side of the satellite because sputtering occurs at significantly lower latitudes than on the ram side (Figure 3 and section III.1). As soon as a $H_2O$ molecule reaches the nightside, it sticks efficiently so that the $H_2O$ exosphere is essentially confined on the dayside. In this scenario, the exosphere at the equator is dominated by the $H_2O$.

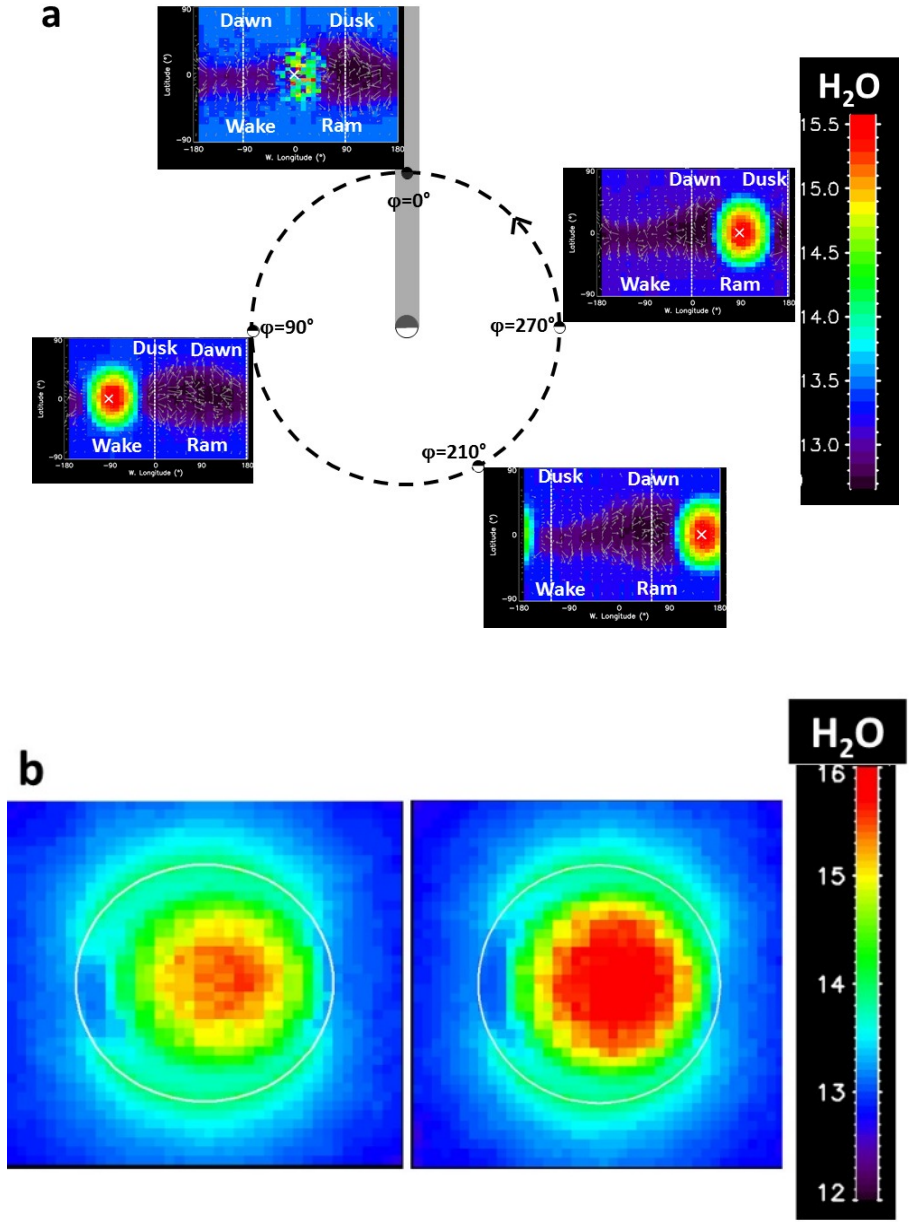

*Figure 8: Panel a: Vertical column densities ($\log_{10}$ cm$^{-2}$) of the $H_2O$ exosphere at different Ganymede phase angles (90°, 210°, 270° and 0°) with respect to latitude (vertical axis) and west longitude (horizontal axis) in the high sublimation scenario. The subsolar point is indicated by the white cross on each panel, the dusk and dawn terminators by vertical dashed black or white lines. The ram and wake hemispheric centers are also indicated. Ganymede's direction of rotation is indicated by the counter-clockwise arrow in the plane of Jovian co-rotation. Jupiter is at the center of each figure. Panel b: Reconstructed column densities (in $\log_{10}$ cm$^{-2}$) of the $H_2O$ exosphere at a phase angle of 90° (left panel) and 270° (right panel) as seen from the Sun.*

For a large sublimation rate, the typical LOS $H_2O$ column density (as seen from the Sun) would peak around the subsolar region with values between ~3.1 and ~8.5×10$^{15}$ $H_2O$/cm² (Figure 8b). The $H_2O$ sputtered from polar regions is also apparent in the Figure 8b. The slight asymmetry of the column density with respect to the subsolar point in Figure 8b is due to the maximum surface temperature occurring in the afternoon. Since dissociation of $H_2O$ is a minor source of H, the column density of H is very close (typically $4\times10^{10}$ H/cm²) to that for the low sublimation rate case (in Figure 8b).

## IV Discussion and Comparison with observations

Barth et al. (1997) reported a H column density of $9.21\times10^{12}$ cm$^{-2}$ in Ganymede's exosphere from the measured Lyman α emission of $0.56\pm0.2\times10^3$ Rayleigh measured by Galileo during its G1 flyby (at a phase angle of 180° and a solar zenith angle of 60° with a latitude of 21°). Such emission intensity was later confirmed with HST (Feldman et al. 2000). The simulated H column density at that phase angle is on the order of $6\times10^{10}$ H/cm$^2$ in the polar region and is extended down to latitudes of 20° because of the particular shape of the open field line boundary at that particular phase angle. Our calculated column density is therefore two orders of magnitude lower than the estimated column density by Barth et al. (1997). A two order of magnitude error in the efficiency sputtering H atoms is difficult to imagine based on measured sputtering yields (Johnson et al. 200**9**).

In combination with resonant scattering, $H_2O$ dissociated via electron impact could be an additional channel producing Lyman α emission. Considering an electron with an energy near 20 eV, the typical cross section for the production of Lyman α by electron impact on $H_2O$ is $0.43\times10^{-18}$ cm$^2$ (Makarov et al. 2004). While trace ~keV hot electrons are a negligible contribution, an accurate knowledge of the core electron energy is particularly important since this cross section increases by one order of magnitude at 50 eV and peaks at $7\times10^{-18}$ cm$^2$ at 100 eV. Using the cross section at 20 eV, the typical column density of $10^{14}$ $H_2O$/cm$^2$ in the low sublimation scenario (Figure 6) and $10^{16}$ $H_2O$/cm$^2$ in the Fray and Schmitt (2009) sublimation scenario would lead to an emission of few Rayleigh and few hundred Rayleigh respectively. The Fray and Schmitt (2009) sublimation scenario would therefore be in rough agreement with Galileo-observed Lyman α emission. The low sublimation scenario could fit the observation for 40 eV core electron temperatures if the sputtering yield was ten times larger. A further potential source of Lyman alpha emission from Ganymede's exosphere is electron impact dissociation of $H_2$, also suggested by Marconi (2007). Ajello et al. (1995) published a cross section for the production of Lyman alpha emission from electron impact dissociation of $H_2$ peaking at $7\times10^{-18}$ $cm^2$ at 60 eV and equivalent to $5\times10^{-18}$ $cm^2$ at an electron energy of 20 eV. Using the typical $2\times10^{14}$ $H_2/cm^2$ column density we simulated, this process would contribute less than a few tens of Rayleigh.

The typical column density of the $O_2$ exospheric component has been reported by Hall et al. (1998) as being between $2.4\times10^{14}$ and $1.4\times10^{15}$ $O_2/cm²$ when observing Ganymede's trailing hemisphere in June 1996 (phase angle around 270°) and by Feldman et al. (2000) as being equal to 0.3 to $5\times10^{14}$ (October 30, 1998 with a phase angle between 290° and 300°). McGrath et al. (2013) also showed the spatial distribution of the auroral emission at different orbital positions characterized by localized bright regions with peak brightnesses between 100 to 400 Rayleigh in good agreement with the previous reported observations. As stated in Section II.4, the $O_2$ sputtering yield was tuned in order to reproduce the reported column density. We therefore simulated column densities varying from 5 to $6.5\times10^{14}$ $O_2/cm²$ averaged over the disk as observed from the Sun, and peaking to values of $2\times10^{15}$ $O_2/cm²$ at the equatorial limb (Figure 6a).

We also reconstructed the auroral emission associated with our simulated column density (Figure 9), by considering electron excitation by a uniform mono-energetic electron population impacting the sputtered $O_2$ in the Alfven wings of Ganymede's magnetosphere (open field line regions displayed in Figure 3) as the primary emission mechanism. We did not attempt to reproduce the ring shape emission in Feldman et al. (2000), McGrath et al. (2013), and Saur et al. (2015). Furthermore, we did not consider an accurate description of the electric currents in the open field line regions as in Payan et al. (2015). The electron population used to calculate the emission brightness is the same as earlier: an electron number density of 70 cm$^{-3}$ and average energy of 20 eV. The emission rate coefficient for the production of OI ($^5$S$^o$) atoms from $O_2$ emitting photons at 135.6 nm wavelength is set to be $1.1\times10^{-9}$ cm$^{-3}$ s$^{-1}$ as in Hall et al. (1998). The 135.6 nm emission produced by direct excitation of the O atoms by electron impact is also included, but is much smaller than the $O_2$ source, in good agreement with Hall et al. (1998). The goal of our calculation is to look for any potential signature in the observed auroral emission induced by the dusk/dawn asymmetry of the $O_2$ exosphere (Figure 4). As shown in Figure 9, the simulated emission intensities vary between 10 and 200 Rayleigh in good agreement with the observed values. More interestingly, at a phase angle of 90° (panel a), Ganymede's dusk exosphere is simulated as being up to two times denser than at dawn consistent with the column density of the $O_2$ exosphere as seen from the Sun being 2.3 times larger on moon's esatern/dusk hemisphere with respect to the column density on the western/dawn hemisphere. This dusk/dawn asymmetry is visible in the simulated emission brightness, the dusk limb emission at mid-latitudes being significantly larger than the dawn limb emission. Remarkably, a similar asymmetry between the dawn and dusk limbs was reported by HST (panel a of Figure 2 in McGrath et al. 2013). At a phase angle of 270° (Figure 9, panel b), the simulated dawn/dusk asymmetry is much smaller (the west/dawn to east/dusk hemispheric average column density ratio begin equal to 1.3), and does not induce a significant emission asymmetry (Figure 9b), as is also seen with HST observations (panel c of Figure 2 in McGrath et al. 2013). As indicated by HST observations near the trailing hemisphere (Fig. 9b) the auroral emissions occur at higher latitudes than at leading (Fig. 9a) since we are facing the ram hemisphere where the open/closed field lines boundary is also shifted to higher latitudes (Fig. 3). The third position reported in McGrath et al. (2013) at a phase angle of 350° (their Figure 2b; near eclipse) is characterized by two bright spots on the right limb. Our simulated image only marginally reproduces such an intense asymmetry between right and left hemispheres (the right to left column densities ratio being equal to 1.2). However, this observation was partially reproduced by Payan et al. (2015) in their Figure 1 using a homogeneous exosphere coupled to a 3D MHD simulation of Ganymede's magnetosphere, suggesting that these bright spots at the limb might be partly due to the current system in Ganymede's magnetospheric Alfven wings. Payan et al. (2015) provided a similar comparison for the two other observed positions (Figures 9a and b) and noted that the bright spot at a phase angle of 105-108° (Figure 9, panel b) could not be reproduced. We suggest here, that this bright spot may be a signature of the strong dusk/dawn asymmetry at the leading hemisphere/plasma wake. Additionally, this dusk/dawn asymmetry could contribute to the almost systematic right to left brightness asymmetry observed in the auroral emissions (McGrath et al. 2013; Saur et al. 2015).

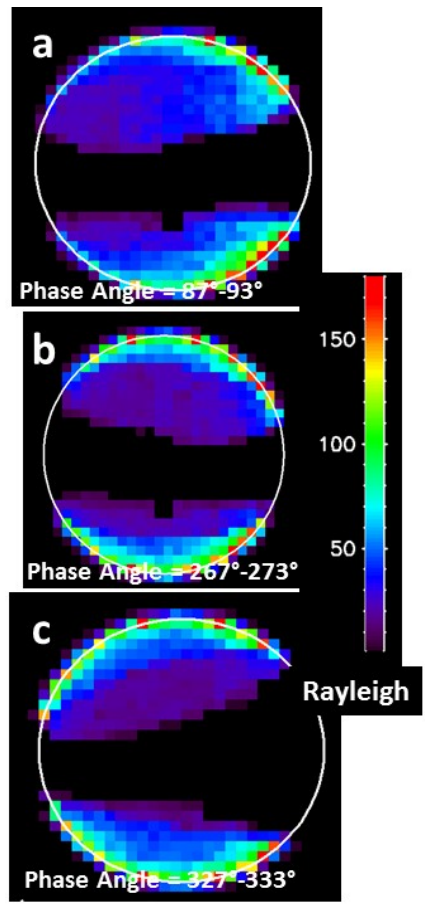

*Figure 9: Simulated images of the HST observed emission brightness reported in McGrath et al. (2013) at three different phase angles as indicated on each panel with emission brightness calculated as explained in the text.*

Using the emission rate coefficient for the 130.4 nm emission, the typical ratio between 135.6 nm and 130.4 nm is equal to 1.8 with decreasing values at the edge of the emission region where the O/$O_2$ relative abundance tends to increase. Hall et al. (1998) and Feldman et al. (2000) reported ratios between 1.2 and 3.2.

Teolis et al. (2016) suggested that the radiolysis efficiency to sputter $O_2$ molecules from an icy surface might be between 50-to-300 times less efficient than predicted by laboratory experiments (Teolis et al. 2010) or as calculated by Cassidy et al. (2013). In this paper, we scaled this efficiency to reproduce the observed $O_2$ column density and found that the ejected $O_2$ rate should be around $3.4\times10^{26}$ $O_2$/s, assuming a flux of Jovian ions precipitating on Ganymede's surface between 3.3 and $3.7\times10^{24}$ particles/s for the thermal component (Leclercq et al. 2016) and $\sim10^{24}$ particles/s for the high-energy component (Fatemi et al. 2016), leading to an average sputtering yield of ~100. Teolis et al. (2016) estimated the yield for ion sputtering on the icy satellites Rhea or Dione to be between 0.01 and 0.03, 4 orders of magnitude less than the yields inferred in this paper. The temperature difference between Rhea, Dione, and Ganymede could only account for a factor of 10 larger. Since an ion precipitation rate that is two orders of magnitude larger is not realistic, we note that the Jovian plasma is more energetic with respect to the Saturnian plasma (Teolis et al. 2016). Such a difference in energy could also lead to an additional order of magnitude larger efficiency at Ganymede than at Dione or Rhea, leaving a difference of two orders of magnitude with respect to the conclusions of Teolis et al. (2016). This difference is re-enforced by noting that two orders of magnitude smaller $O_2$ column density would place atomic products below the detection limit of FUV oxygen aurorae observations by HST. Teolis et al. (2016) also suggested that for reproducing Rhea's exosphere evolution, it was needed to introduce the notion of a shallow 1-cm regolith allowing for the diffusion of $O_2$ into the regolith and its shielding from the impacting ions. The release of the $O_2$ is then allowed when the regolith rotates into sunlight and induces the diffusion of the $O_2$ molecules back to the surface where they can be released by sputtering. Applied to Ganymede's case, such a process would increase the reservoir of $O_2$ available for ejection (by desorption or sputtering) and therefore help reconcile Teolis et al. (2016) sputtering efficiency and the efficiency inferred by our modelling. It would imply that more than 99% of the $O_2$ originally produced by radiolysis accumulates in the regolith over time and is released into the exosphere by thermal desorption and/or direct sputtering. In such a scenario, the $O_2$ exosphere would be much more extended in longitude and latitude. We should also expect a much less asymmetric dawn to dusk terminator, in which the $O_2$ exosphere behaves like the $H_2O$ exosphere. Moreover, given that $O_2$ is more efficiently adsorbed and trapped inside the regolith than simulated here, a dawn enhancement would be favored by the release of trapped $O_2$ at sunrise and a decrease in reservoir content with increasing solar zenith angle. $O_2$ trapping in the regolith has actually been suggested by ground-based observations peaking at the trailing hemisphere (Spencer et al. 1995), and confined at the equatorial regions (Calvin and Spencer 1997). It was shown by Johnson and Jesser (1997) that $O_2$ (and $O_3$) bubbles can be trapped as voids in Ganymede's regolith with the caveat that they can be destroyed should the incident ion flux be too high. The absence of $O_2$ at the poles, as well as the significantly less $O_2$ at the leading hemisphere/plasma wake might be explained by a more efficient eroding of the trapped $O_2$ bubbles by the incident plasma.

**V Conclusions**

In this paper, we present the first model of Ganymede's exosphere taking into account Jupiter's gravitational influence and the evolution of Ganymede's exosphere along its orbit around Jupiter. This 3D Monte Carlo model takes into account sublimation and sputtering as the main processes of ejection, electron impact ionization and dissociation in the exosphere as well as a detailed interaction between the exosphere and the surface. The role of Ganymede's magnetosphere is described in a simple way which allows us to clearly distinguish signatures associated with the magnetosphere to those associated with the formation and evolution of the exosphere. Ganymede's exosphere appears to be highly stratified, displaying significant structure in terms of density and composition due to the variation of the ejection mechanisms and molecular movement over its orbital period. In particular, our results suggest that the shape of Ganymede's $O_2$ exosphere at a given orbital position is highly dependent on the production earlier in its orbit. We also showed that the $H_2O$ exosphere is very different for the high sublimation rate and a low sublimation rate models. Moreover, even if Ganymede's atmosphere is quasi-collisional, collisions are shown to only have a minor effect on the exospheric density distribution.

In this work, we show that, by observing Ganymede's multi-species, water product exosphere along its orbit, significant information on the way Ganymede interacts with its environment should be accessible:

- The formation of the exosphere is a complex process driven by the impacting plasma and the resulting space weathering of the surface, whose efficiency depends on the surface characteristics (e.g. temperature, composition, structure) and on the Jovian plasma parameters (e.g. flux, energy and spatial distribution at the surface). The spatial distribution for the lighter species of the sputtered exosphere should directly reflect the spatial characteristics of the interaction between Ganymede and Jupiter's magnetosphere. Moreover, the sputtered exospheric temporal variability and composition should be highly dependent on the surface temperature and composition and could therefore be used to constrain these two surface properties.

- The global 3-D spatial distribution of the exospheric density will depend on the influence of Jupiter and Ganymede's gravitational fields. The fate of the ejecta is also species-dependent due to the species dependent surface interactions. As an example, the dusk/dawn asymmetry in $O_2$ highlighted in this paper depends on the mobility across Ganymede's surface affected by absorption, accumulation and re-emission from the regolith. The observed spectral emission signatures, such as the auroral emissions, could be potentially used to characterize not only the Jovian-Ganymede interaction (Saur et al. 2015), but also the spatial distribution of the exosphere.

- Ultimately, tracking the evolution of Ganymede's exospheric morphology and content along its orbit should help constrain its atmospheric formation, and the interaction of its magnetospheric plasma with the atmosphere and surface. The Galileo spacecraft showed the dichotomy in the surface albedo between the trailing and leading hemispheres, which suggest that, Ganymede's exospheric source can depend strongly on geographic longitude, an effect we did not consider in this work, but will implement in the future. To better constrain the full structure of the exosphere we suggest full orbital observations of Ganymede's FUV oxygen aurorae capable of indicating the evolution of Ganymede's atmosphere.

**Acknowledgement:** This work is part of HELIOSARES Project supported by the ANR (ANR-09-BLAN-0223) and ANR MARMITE-CNRS (ANR-13-BS05-0012-02). Authors are also indebted to the "Système Solaire" program of the French Space Agency CNES for its support. Authors also acknowledge the support of the IPSL data center CICLAD for providing us access to their computing resources and data. Data may be obtained upon request from F. Leblanc (email:francois.leblanc@latmos.ipsl.fr).